\documentclass{ceab}   

\usepackage{epsfig}     
\usepackage{graphicx}   

\usepackage{ceabbib}     
\usepackage[T1]{fontenc}
\begin{document}

\def\tit{A Mysterious System FF\,Cam}
\def\aut{Garrel et al.}
\def\str{67--72}

\title{Spectroscopy of the mysterious Be system FF\,Cam}

\author{T. GARREL$^1$, A.~S. MIROSHNICHENKO$^2$, S. DANFORD$^2$,\\
S. CHARBONNEL$^3$, F. HOUPERT$^4$, K.~N. GRANKIN$^5$, and A.~V. KUSAKIN$^6$\\
\\
\it $^1$ Observatoire de Foncaude, Montpellier, France\\
\it $^2$ University of North Carolina at Greensboro,\\
\it Department of Physics and Astronomy, Greensboro, NC, USA\\
\it $^3$ Durtal Observatory, Durtal, France\\
\it $^4$ Verny Observatory, Verny, France\\
\it $^5$ Crimean Astrophysical Observatory, Nauchny, Ukraine\\
\it $^6$ Fessenkov Astrophysical Institute, Almaty, Kazakhstan}

\maketitle

\begin{abstract}
FF\,Cam is a variable star near the North celestial pole with
hydrogen lines in emission. Its optical variability of $\sim$0.3 mag
was discovered by HIPPARCOS. The spectral type assigned to the star
in SIMBAD is B9, but its position coincides with a ROSAT X--ray
source. This suggests the presence of a high-temperature region in
the system that could originate at or near a companion object. We
undertook a spectroscopic monitoring of FF\,Cam since the beginning
of 2012 and found an extremely variable H$\alpha$ line profile as
well as periodically variable radial velocities of numerous
absorption lines. The main conclusion from our study is that FF\,Cam
is a binary system with an orbital period of 7.785 days, a B--type
primary and a K--type secondary component. We discuss the spectral
features, their variations, and the nature of FF\,Cam.\end{abstract}


\section{Introduction}

FF\,Cam = HD\,60062 is a fairly bright ($V \sim$ 7.5--8.0 mag) star
not far from the North Celestial Pole (R.A. 7h 47m, Dec.
+81$^{\circ} 40^{\prime}$, 2000). It was discovered as a variable
star by the HIPPARCOS mission (ESA 1997) that obtained 146
measurements of its visual magnitude on 38 different days in
1989--1993 (see Fig. \ref{f3}b). These data were first analyzed by
Woitas (1997), who found no periodic variations. The object was
included in the General Catalog of Variable Stars as FF\,Cam
(Kazarovets et al.\,1999) and classified as a Be star. We are still
searching for the origin of this classification, although it does
have an emission-line spectrum. An ROSAT X--ray source with a flux
of 0.15$\pm$0.02 counts\,s$^{-1}$ was recently found in
0$.\!\!^{\prime\prime}$1 from the visible/IR star position
(Haakonsen \& Rutledge 2009). FF\,Cam is located far from the
Galactic plane (b = 29$^{\circ}$) that implies a low reddening. Its
HIPPARCOS parallax leads a distance of 580$^{+270}_{-150}$ pc (ESA
1997) which may be wrong, if the object is a binary system (see
Sect. \ref{results}). Both the distance and apparent brightness
suggest that the system is not very luminous. The current
presentation is the first study of the spectroscopic behavior and
the spectral energy distribution of FF\,Cam.

\begin{figure}[htb]
\begin{center}
\begin{tabular}{cc}
\includegraphics[width=5.4cm,height=4.9cm]{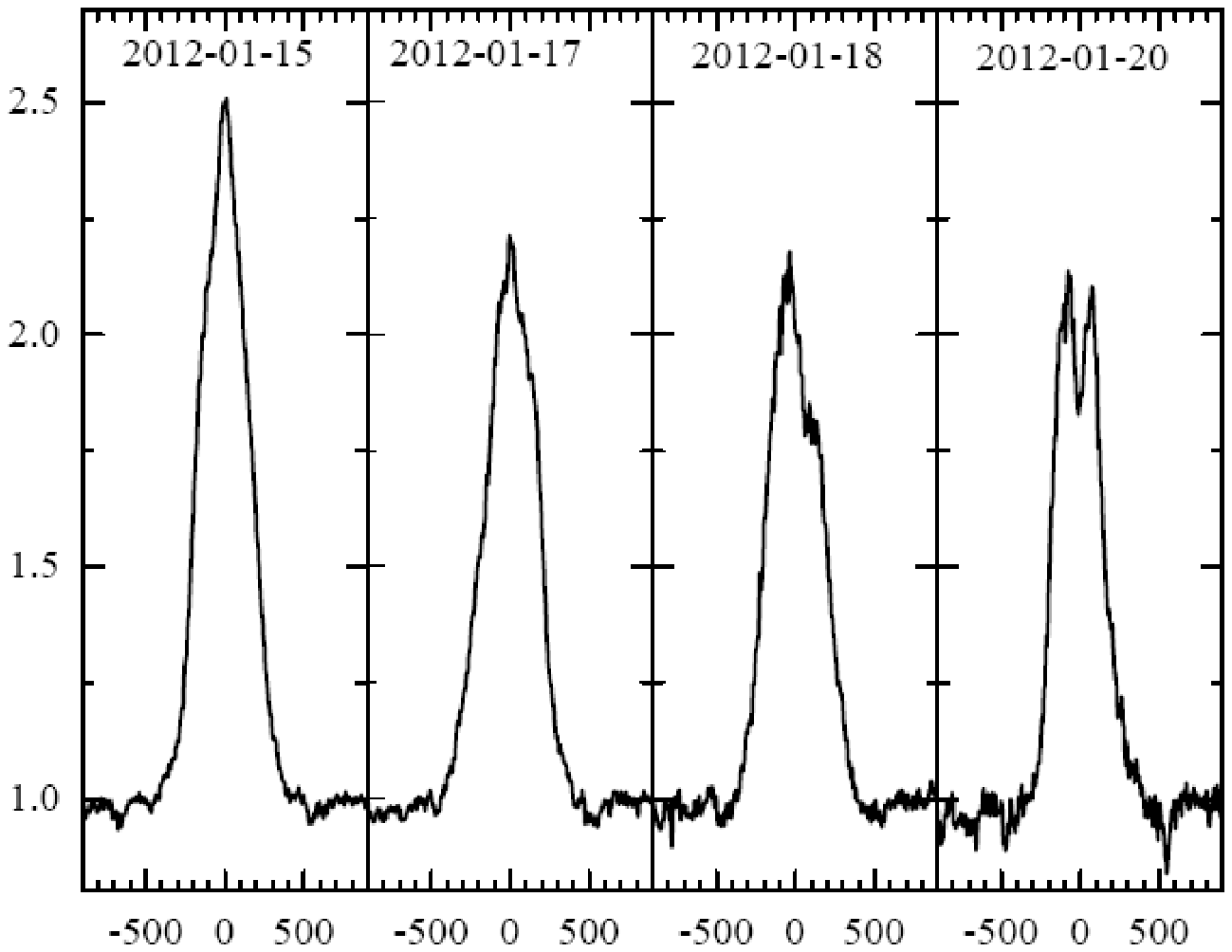}&
\includegraphics[width=5.4cm,height=4.8cm]{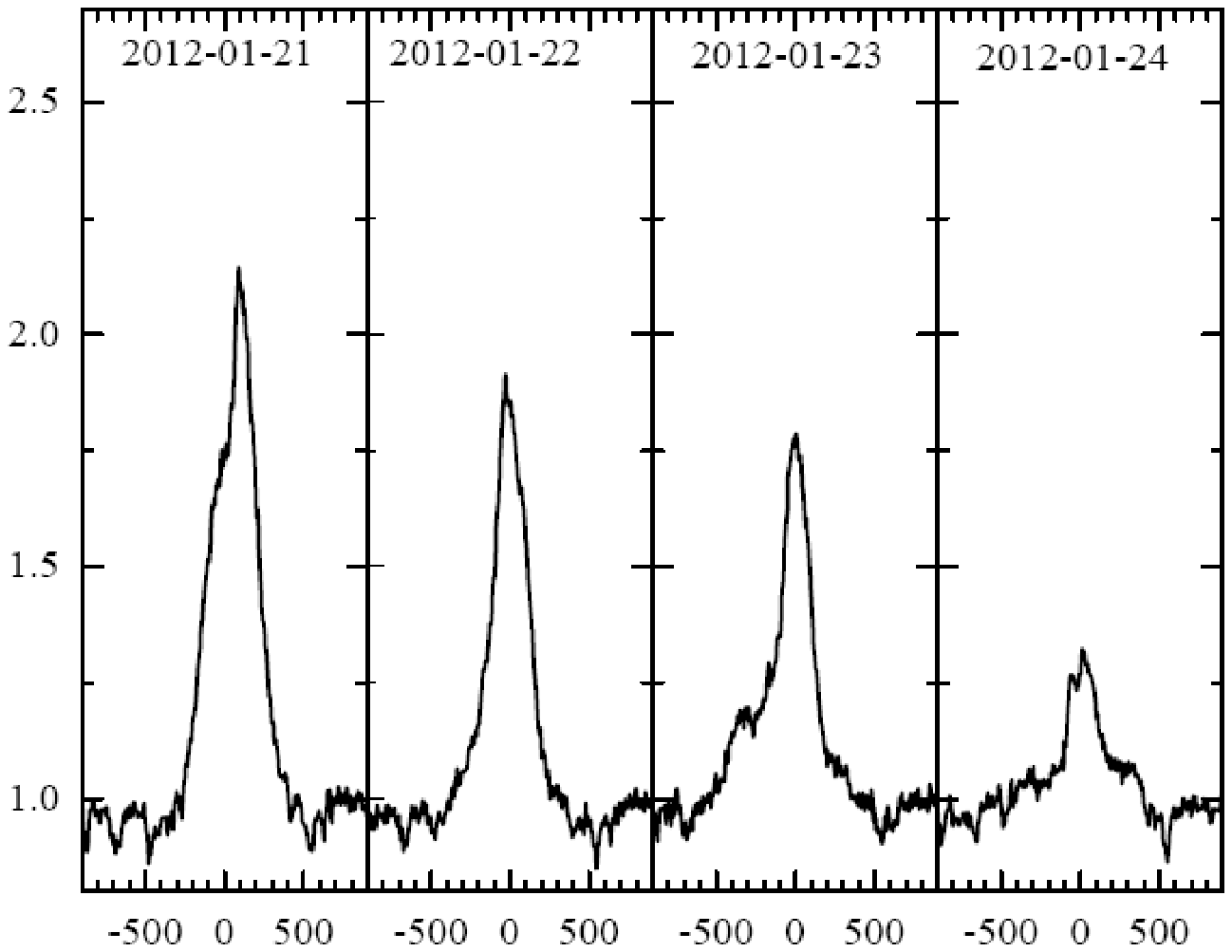}\\
\includegraphics[width=5.4cm,height=4.8cm]{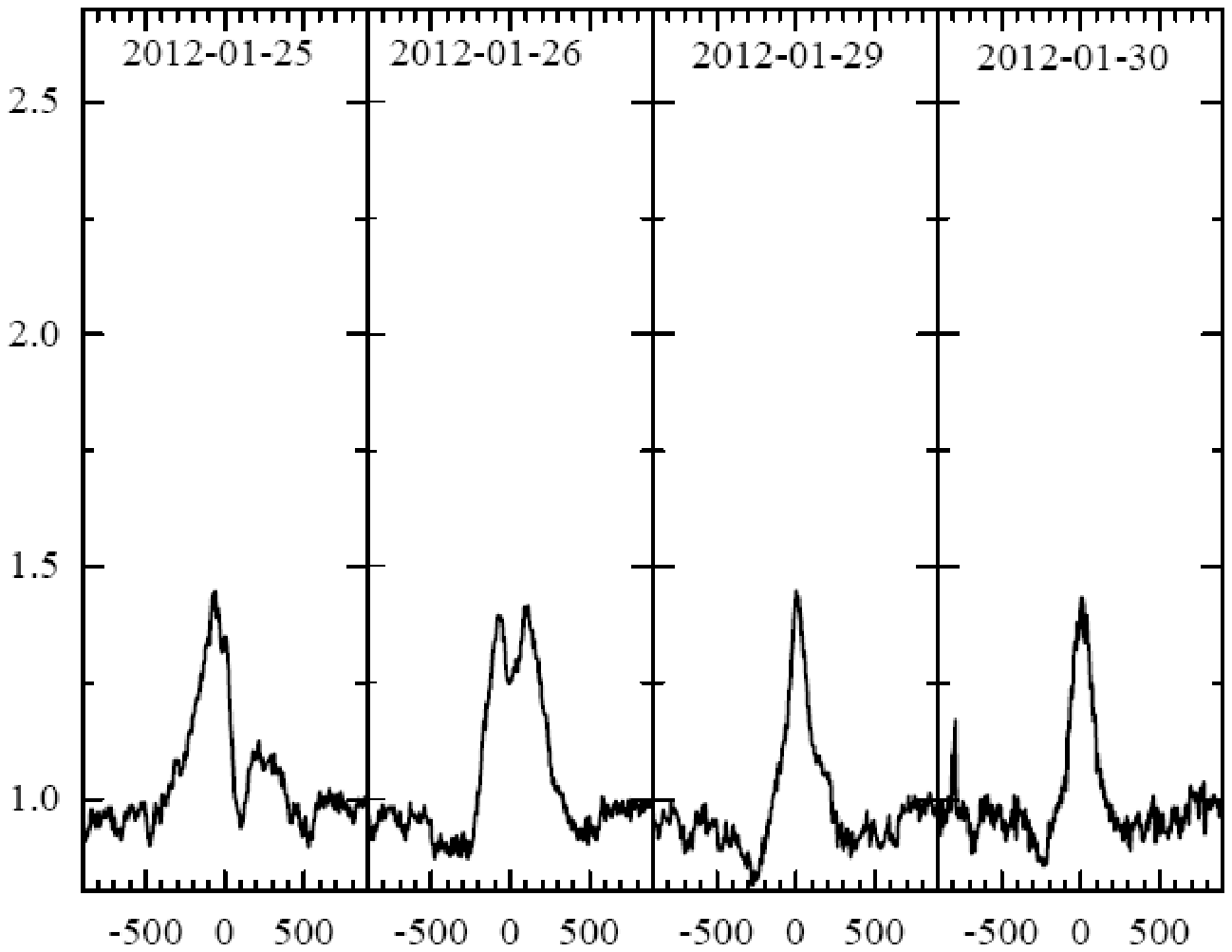}&
\includegraphics[width=5.4cm,height=4.8cm]{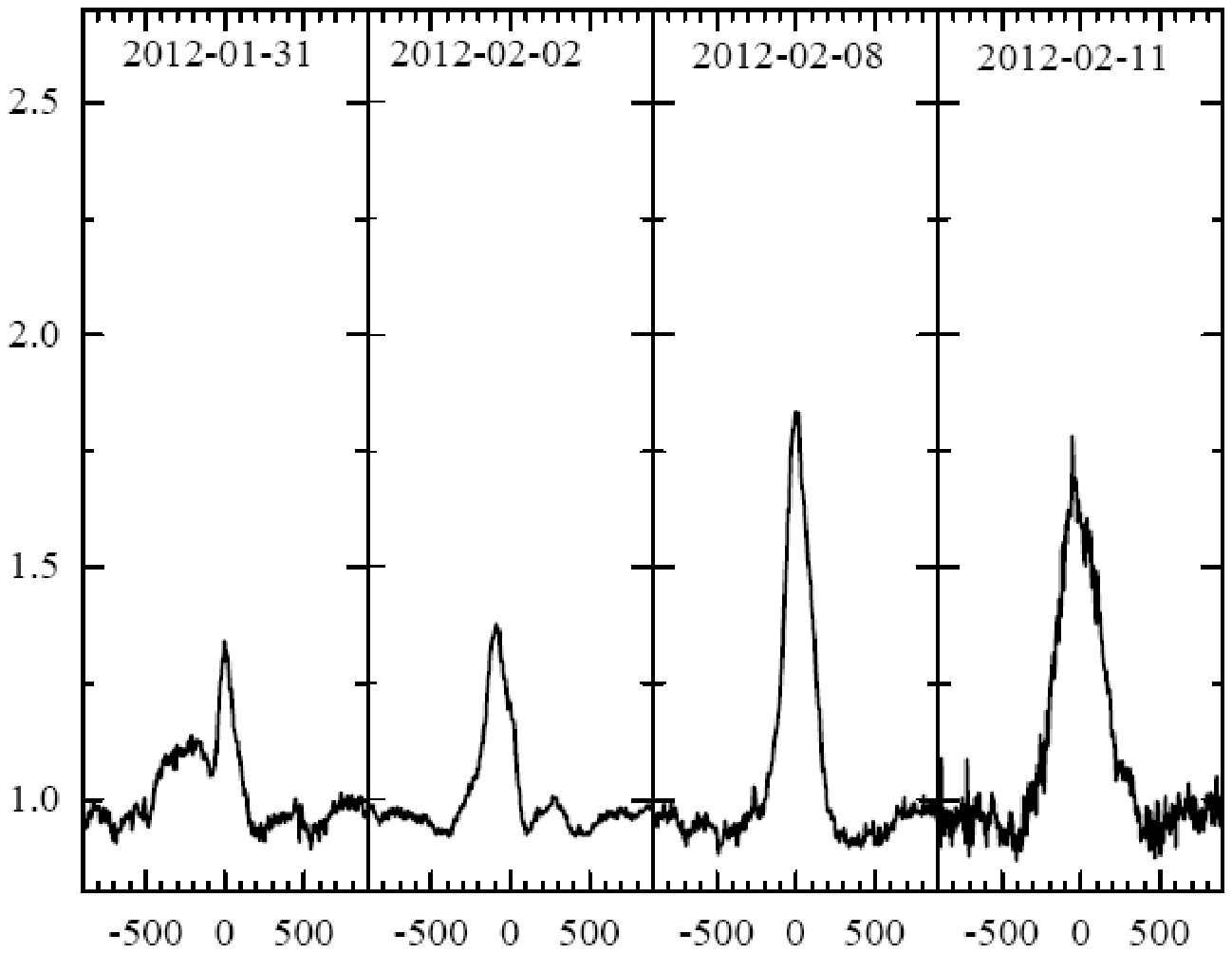}\\
\end{tabular}
\end{center}
\caption{The H$\alpha$ line profile variations in the spectrum of
FF\,Cam in January -- February 2012. Intensity is normalized to the
underlying continuum, radial velocity is heliocentric. } \label{f1}
\end{figure}

\section{Observations}

Photometric $UBVRI$ observations were obtained on 2012 April 13 and
15 with the 1.25--m telescope of the Crimean Observatory in Ukraine.
$BVR$ observations were obtained on 2012 April 11, 12 and May 24
with a 0.2--m Newtonian reflector at the Tien-Shan Observatory near
Almaty, Kazakhstan. Medium-resolution spectra were obtained at three
amateur sites in France (Montpellier, Durtal, and Verny) and in the
USA at the Three College Observatory (TCO, near Greensboro, North
Carolina). Two types of spectrographs were used: long-slit LHires
III in the H$\alpha$ region with a spectral resolving power $R \sim$
17,000 and \'echelle ($\lambda\lambda$ 4300--7200 \AA, $R \sim$
10,000). All the spectrographs were manufactured by the Shelyak
company (www.shelyak.com). In total we obtained over 100 spectra in
2010--2012, mostly in January--April 2012. IRAF was used to reduce
the TCO data, the amateurs data were reduced with software packages
developed for amateur spectrographs
(Audela\footnote{http://www.audela.org/dokuwiki/doku.php/en/start}
and
ISIS\footnote{http://www.astrosurf.com/buil/isis/isis$\_$en.htm}.)

\section{Results} \label{results}

\begin{figure}[htb]
\begin{center}
\begin{tabular}{cc}
\includegraphics[width=5.7cm,height=5.8cm]{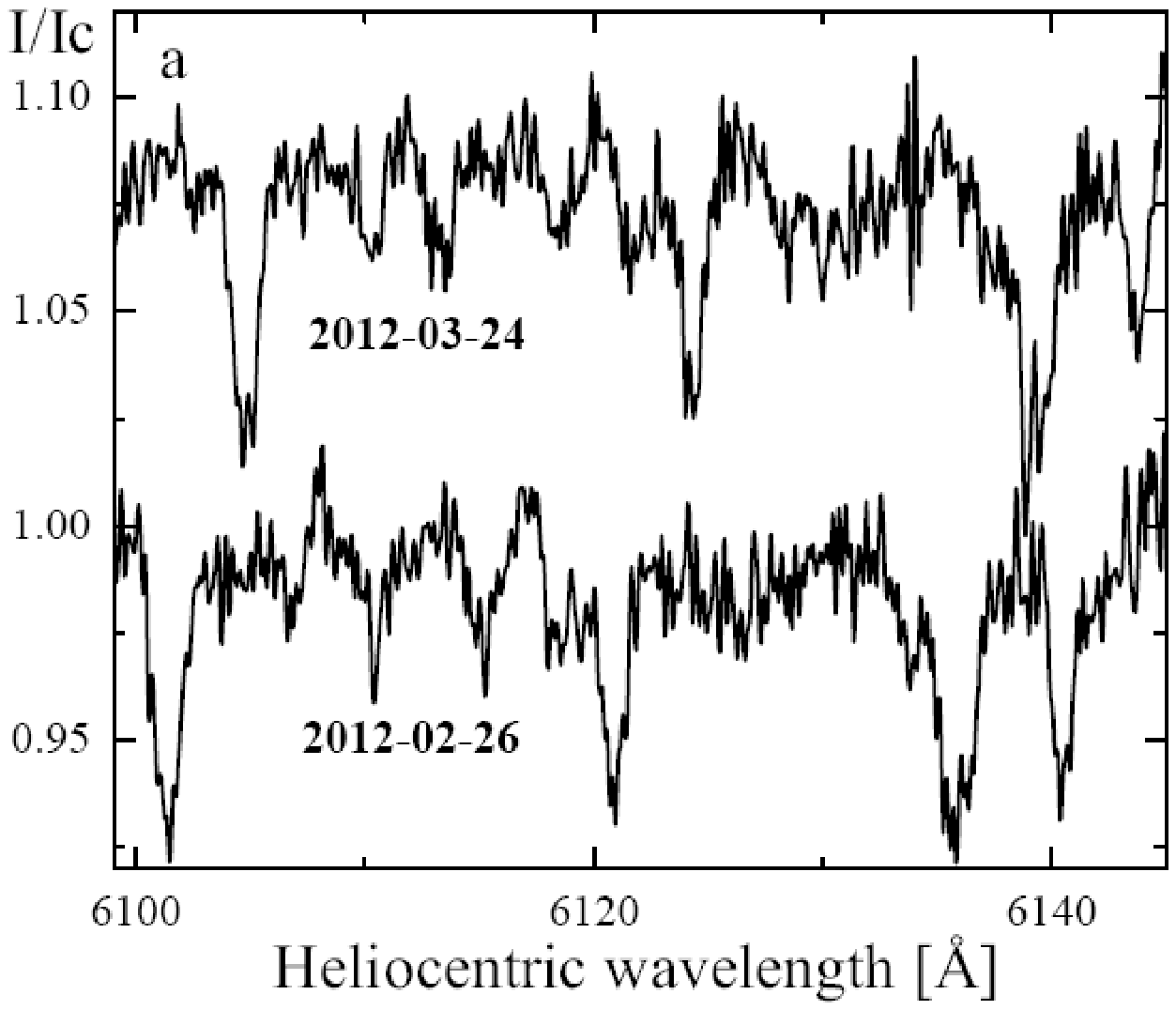}&
\includegraphics[width=5.7cm,height=5.7cm]{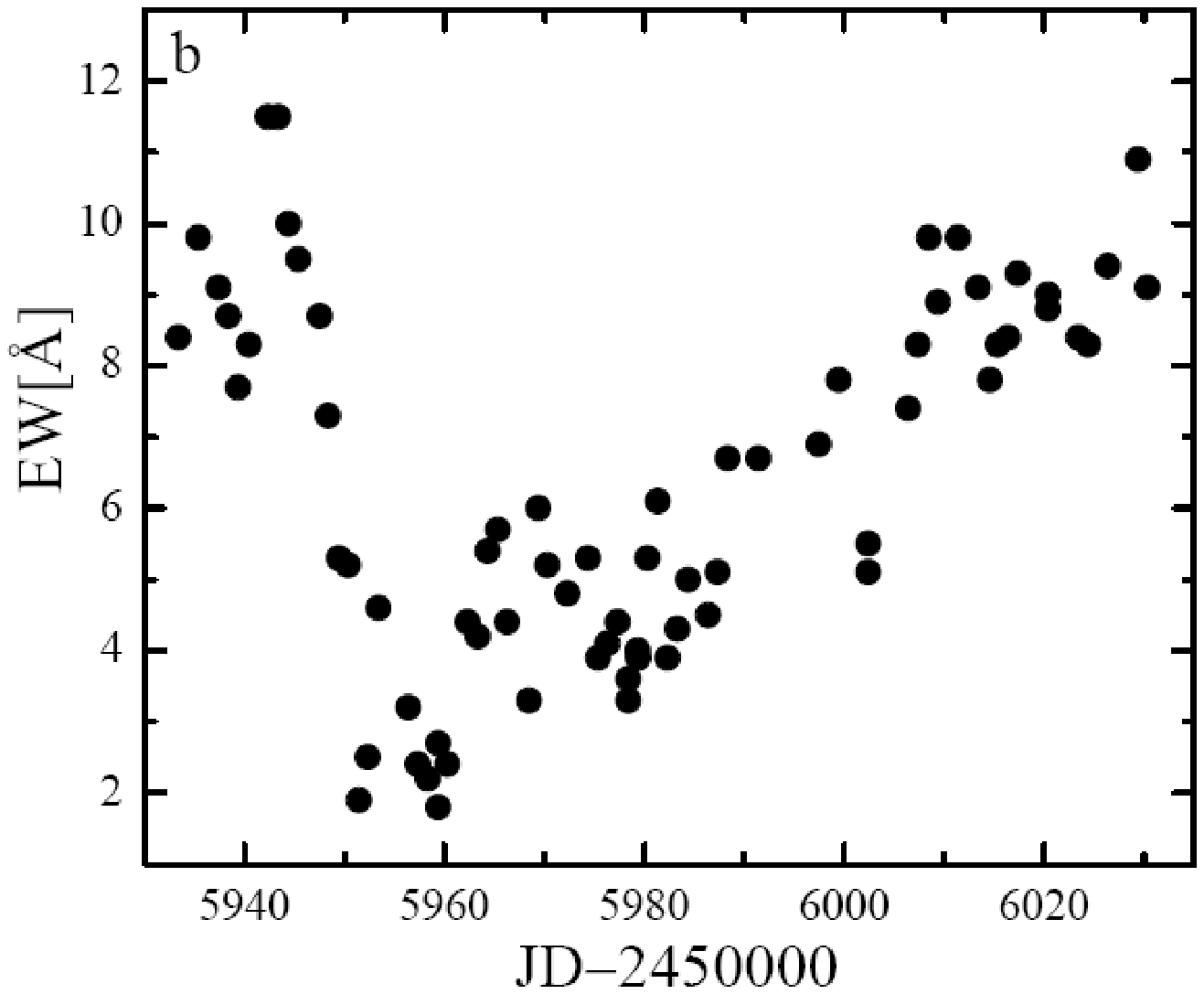}\\
\end{tabular}
\end{center}
\caption{{\bf Panel a.} Absorption lines in the spectrum of FF\,Cam
at different phases of the orbital cycle. Intensity is normalized to
the underlying continuum. Radial velocity is measured by fitting the
line profiles to a Gaussian. {\bf Panel b.} Variations of the
H$\alpha$ line equivalent width measured by intensity integration
with the line profile in the continuum normalized spectra. Only the
emission component of the line profile was used for the
measurements.} \label{f2}
\end{figure}

The main features of the FF\,Cam optical spectrum are the following.
The Balmer H$\alpha$ and H$\beta$ lines are seen in variable
emission with mostly single-- or double--peaked profiles (see Fig.
\ref{f1}), while almost no emission is observed in H$\gamma$. Fe
{\sc ii} lines show weak double-peaked emission profiles. He {\sc i}
lines are in absorption. Numerous absorption lines of neutral metals
are weak ($\le$ 10\% of the continuum, see Fig. \ref{f2}a).

Strong absorption components of the H$\beta$ and H$\gamma$ lines and
strong pure absorption He {\sc i} lines along with weak absorption
lines of neutral metals imply that FF\,Cam has a composite spectrum.
It is a binary system with a brighter B--type component (later than
B2) and a cooler, fainter component probably of an early K--type.
The temporal behaviour of the emission-line spectrum (Fig.
\ref{f2}b) implies that the amount of circumstellar gas in the
system varies with time.

\begin{figure}[htb]
\begin{center}
\begin{tabular}{cc}
\includegraphics[width=5.5cm,height=5.0cm]{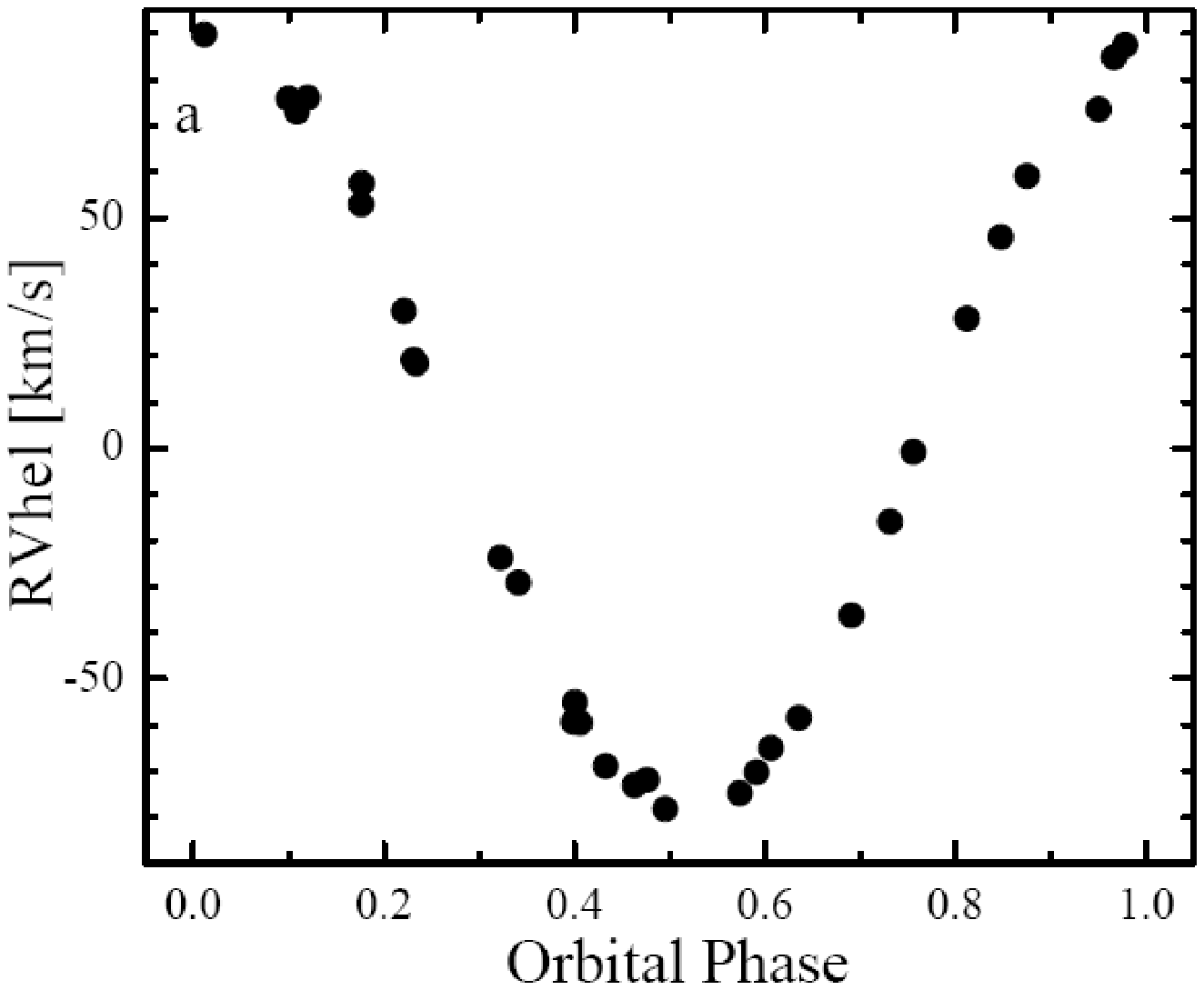}&
\includegraphics[width=5.5cm,height=5.0cm]{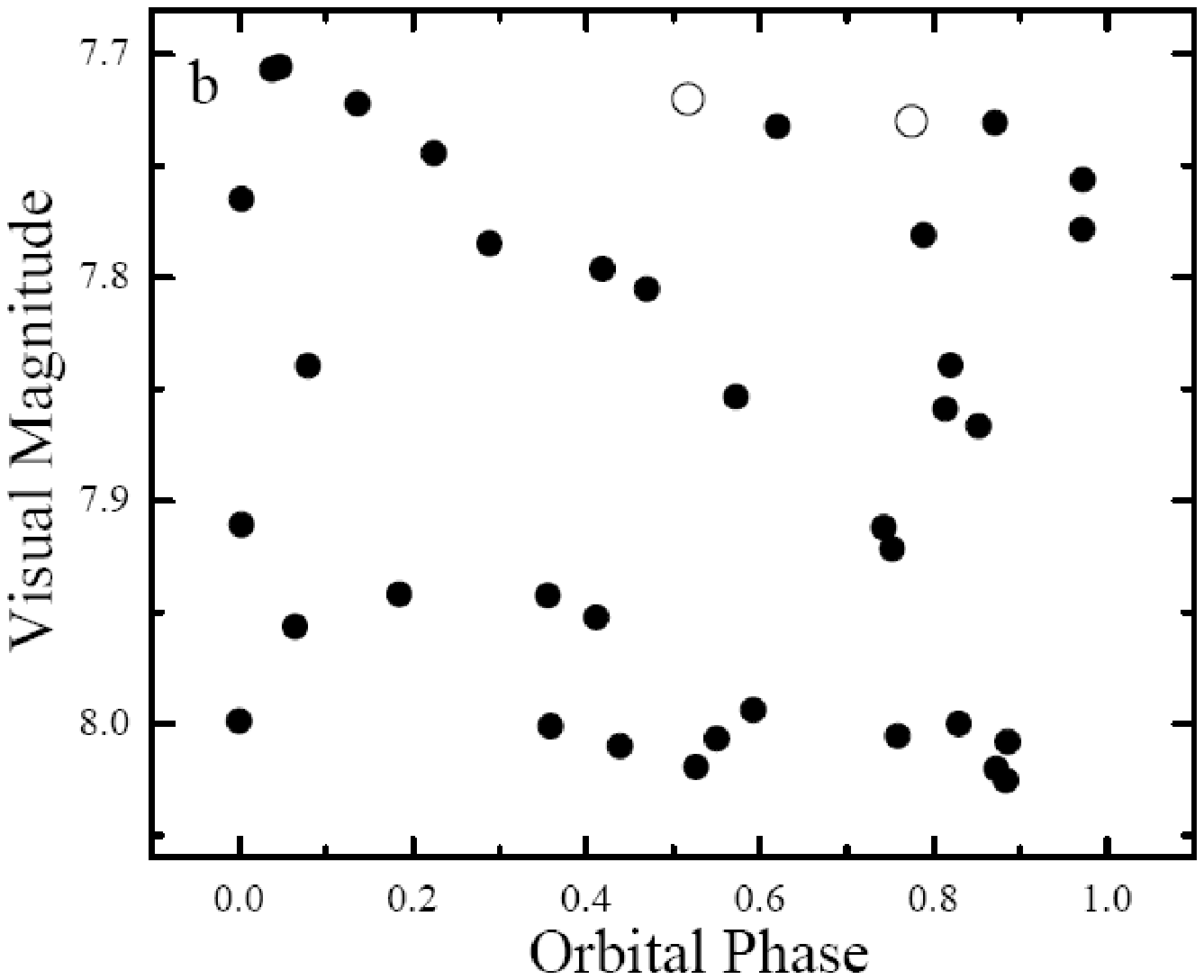}\\
\end{tabular}
\end{center}
\caption{Panel a. Absorption-line radial velocity curve of FF Cam.
Panel b. Visual brightness variations of FF\,Cam plotted against the
spectroscopic orbital phase. Filled circles are HIPPARCOS data, open
circles are $V$--band magnitudes from Crimea.} \label{f3}
\end{figure}

Analyzing radial velocity variations, we found them strictly
periodic and sinusoidal (Fig. \ref{f3}a). Therefore, the binary
orbit is circular. The following orbital elements were determined.
Radial Velocity Maxima = JD 2455941.594 + 7.785*E (days). The radial
velocity semi-amplitude for the cool component K$_{2}$ = 85.0
km\,s$^{-1}$. These elements lead to a mass function of f(M) = 0.5
M$_{\odot}$.

The emission line intensity variations do not correlate with the
orbital period. Since the stars are very close to each other, the
cool component most likely fills its Roche lobe. This can cause mass
transfer into the Roche lobe of the hot component and formation of
an accretion disk around it.

Absorption lines of the hot component move roughly in anti-phase
with those of the cool component, but they seem to be contaminated
by the accretion disk. Small variations of the cool component line
intensities and chaotic brightness variations (Fig. \ref{f3}b)
suggest no eclipses in the system.

\begin{figure}[htb]
\begin{center}
\includegraphics[width=6.0cm,height=5.5cm]{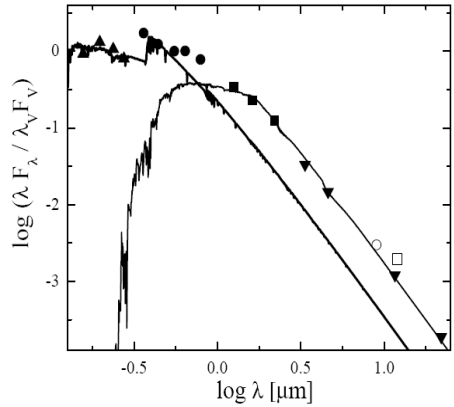}
\end{center}
\caption{Spectral energy distribution of FF Cam. Symbols: filled
upward triangles -- TD1 UV satellite, filled circles -- Crimean
$UBVRI$, filled squares -- 2MASS $JHK$, filled downward triangles --
WISE, open circle -- AKARI, open square -- IRAS. The thick line is a
Kurucz (1993) model atmosphere for a B7--star (70\% contribution to
the observed flux), the thin line is a model atmosphere for a
K0--star (30\% contribution). No interstellar extinction is taken
into account.} \label{f4}
\end{figure}

The low value of the mass function and the presence of a B--type
star in the system imply that the system is viewed nearly pole-on.
For a lower limit of 5 M$_{\odot}$ for the system mass, the orbital
inclination angle is 27$^{\circ}$. Assuming the HIPPARCOS distance,
no interstellar reddening, and a components brightness ratio of
$\Delta V$ = 1 mag (Fig. \ref{f4}), the cool component's radius is
10 R$_{\odot}$. The X--ray flux is $\sim 10^{-5}$ of the hot star
optical flux. This is over an order of magnitude larger than the
expected photospheric flux from the B--type component. Our current
data offer no explanation to this fact.

\section{Conclusions}

We have found that FF\,Cam is not a Be star, but rather a
short-period (7.785 days) semi-detached binary system with mass
transfer from the cool component to the hot component. The mass
transfer is variable and results in fast variations of the
emission-line profiles as well as in the observed photometric
variations. High-resolution spectroscopy with a good phase coverage
and high signal-to-noise spectra are needed to constrain the
component spectral types and orbital parameters.

\section*{Acknowledgements}
A.M. acknowledges support from the American Astronomical Society
International Travel Grant program and from the Department of
Physics and Astronomy of the University of North Carolina at
Greensboro. This research has made use of the SIMBAD database,
operated at CDS, Strasbourg, France, data products from the Two
Micron All Sky Survey (2MASS) and the Wide-Field Infrared Survey
Explorer ({\it WISE}), and the BeSS database, operated at LESIA,
Observatoire de Meudon, France (accessible at\\
http://basebe.obspm.fr).

\section*{References}
\begin{itemize}
\small
\itemsep -2pt
\itemindent -20pt
\item[] ESA 1997, European Space Agency SP--1200. The Hipparcos and Tycho
Catalogues.
\item[] Haakonsen, C.~B., and Rutledge, R.~E.: 2009, {\it \apjs} {\bf 184},
138.
\item[] Kazarovets, E.~V., Samus, N.~N., Durlevich, O.~V., et al.: 1999, IBVS,
4659.
\item[] Kurucz, R.L.: 1993, {\it CD-ROM No. 13, Smithsonian Astrophysical Observatory}.
\item[] Woitas, J.: 1997, Inform. Bull. Var. Stars, No 4444.
\end{itemize}
\end{document}